\documentclass{ws-ijbc}
\usepackage{ws-rotating}     

\usepackage{color}

\begin{document}

\catchline{}{}{}{}{} 

\markboth{Boreux et al.}{Control of accelerator maps}

\title{(This paper is for the Special Issue edited by \\
  Prof. Gregoire Nicolis , Prof. Marko Robnik, Dr. Vassilis Rothos and
  Dr. Haris Skokos) \\ Efficient control of accelerator maps}

\author{JEHAN BOREUX} \address{Namur Center for Complex Systems,
  naXys, University of Namur\\ Rempart
  de la Vierge 8, Namur, 5000, Belgium\\
  jehan.boreux@fundp.ac.be}

\author{TIMOTEO CARLETTI} \address{Namur Center for Complex Systems,
  naXys, University of Namur\\ Rempart
  de la Vierge 8, Namur, 5000, Belgium\\
  timoteo.carletti@fundp.ac.be}

\author{CHARALAMPOS SKOKOS} \address{Max Planck Institute for the
  Physics of Complex Systems,\\ N\"{o}thnitzer Str. 38, D-01187,
  Dresden, Germany\\ and\\ Center for Research and Applications of
  Nonlinear Systems,\\ University of Patras, GR-26500, Patras,
  Greece\\ hskokos@pks.mpg.de}

\author{YANNIS PAPAPHILIPPOU}
\address{CERN\\
  CH-1211 Geneva 23, Switzerland \\
  yannis@cern.ch }

\author{MICHEL VITTOT}
\address{Centre de Physique Th\'eorique\\
  CNRS Luminy - case 907 - 13288 Marseille cedex 9, France\\
  vittot@cpt.univ-mrs.fr }

\maketitle

\begin{history}
\received{(to be inserted by publisher)}
\end{history}

\begin{abstract}
  Recently, the Hamiltonian Control Theory was used in \cite{BC2010}
  to increase the dynamic aperture of a ring particle accelerator
  having a localized thin sextupole magnet. In this paper,
  these results are extended by proving that a simplified version of the
  obtained general control term leads to significant improvements of
  the dynamic aperture of the uncontrolled model. In addition,  the dynamics
  of flat beams based on the same accelerator model can be significantly
  improved by a reduced controlled term applied in only 1 degree of freedom.
\end{abstract}

\keywords{Hamiltonian control, Particle accelerators, Dynamical
  aperture, Symplectic maps, SALI method}

\newcommand{\mc}{\mathcal}

\section{Introduction}
\label{sec:intro}

Hamiltonian Control Theory has been developed in
\cite{V2004,CVECP2005} aiming to improve some selected features
of a given Hamiltonian system (like to reduce its chaotic behavior), by
adding a `small control term' which slightly alters the Hamiltonian
of the system.

This technique was adapted in \cite{BC2010} to the case of symplectic
maps. In particular, it was applied to a four-dimensional (4D) map
which models a simplified accelerator ring with sextupole
nonlinearity, succeeding to increase the stability domain around the
nominal circular orbit (the so-called dynamic aperture - DA). This
procedure allowed the construction of a sequence of control terms of
increasing complexity, that successively improve the dynamics of the
original map. Actually, the larger the DA became the functional
complexity of the control term increased, and the required CPU time
for the evolution of orbits grew.  Real accelerators are composed of such simplified maps (see for example \cite{F1998}). Thus, we believe that the study of these maps can guide the research for increasing the DA of more complicated models.

The aim of the present paper is to show that also simplified versions
of the general control term can lead to a significant increase of the
DA. For accelerators the addition of a control term can be interpreted
as the addition of an extra magnet. Since magnetic fields depend only
on spatial variables we investigate the efficiency of a modified
control term obtained by neglecting the dependence on the momenta of
the general control term constructed in \cite{BC2010}.

In many real accelerators the vertical extend of the beam is much
smaller than its horizontal size. Such cases can be approximated by
considering ideal, flat beams of zero height, whose dynamics is
described by a restriction of the general 4D accelerator model to a
2-dimensional subspace. We also apply our control theory to a 2D map
model of this type, improving the stability of the flat beam.

\section{The model of a simplified accelerator ring}
\label{sec:themodel}

As in \cite{BT1991,BS2006,BC2010} we consider a simplified accelerator
ring with linear frequencies (tunes) $q_x$, $q_y$, having a localized
thin sextupole magnet. The evolution of a charged particle is modeled
by the 4D symplectic map
\begin{equation}
  \label{eq:4DFODO}
  \left(
    \begin{smallmatrix}
      x_1^{\prime}\\x_2^{\prime}\\x_3^{\prime}\\x_4^{\prime}
    \end{smallmatrix}
\right)=\left(
  \begin{smallmatrix}
    \cos \omega_1 & -\sin\omega_1 & 0 & 0 \\
    \sin \omega_1 & \cos\omega_1 & 0 & 0 \\
    0 & 0 & \cos \omega_2 & -\sin\omega_2 \\
    0 & 0 & \sin \omega_2 & \cos\omega_2 \\
  \end{smallmatrix}
\right)  \left(
    \begin{smallmatrix}
      x_1\\x_2+x_1^2-x_3^2\\x_3\\x_4-2x_1x_3
    \end{smallmatrix}
\right)=T_{\scriptscriptstyle{S}}\left(
    \begin{smallmatrix}
      x_1\\x_2\\x_3\\x_4
    \end{smallmatrix}
\right)\, .
\end{equation}
where $x_1$ {($x_3$)} denotes the initial deflection from the ideal
circular orbit in the horizontal {(vertical)} direction before the
particle enters the element, and $x_2$ {($x_4$) is the associated
  momentum.} Primes denote positions and momenta after one turn in
the ring.  The parameters $\omega_1$ and $\omega_2$ are related to the
accelerator's tunes $q_x$ and $q_y$ by $\omega_1 = 2\pi q_x$ and
$\omega_2 = 2\pi q_y$.  In our study, we set $q_x=0.61803$ and
$q_y=0.4152$ corresponding to a regular orbit as shown in \cite{VIB1997}\footnote{We remark that in the theory we developed $q_x$ and $q_y$ have been
considered as parameters. So, our results can be straightforwardly
applied to other cases, remaining valid for tunes satisfying  a non-resonant condition (see section \ref{sec:methods} below). We also note that our method can be easily adapted to the resonant case, but  the construction of  a new control term is required.}.  The particle dynamics at the $n$-th turn, can be
described by the sequence
$(x_1{(n)},x_2{(n)},x_3{(n)},x_4{(n)})_{n\geq 0}$, where the
$(n+1)$-th positions and momenta are defined as a function of the
$n$-th ones by~\eqref{eq:4DFODO}.

\section{Theoretical considerations and numerical techniques}
\label{sec:methods}

For sake of completeness, let us briefly recall the theoretical framework
developed in \cite{BC2010}).  Map \eqref{eq:4DFODO} naturally
decomposes in a integrable part, the rotation by angles $\omega_1$,
$\omega_2$ in the planes $x_1,x_2$ and $x_3,x_4$ respectively, and a
quadratic \lq\lq perturbation\rq\rq, and can be obtained as the
time-$1$ flow of the following Hamiltonian systems
\begin{equation}
  \label{eq:hamrot}
  H(x_1,x_2,x_3,x_4)=-\omega_1\frac{x_1^2+x_2^2}{2}-\omega_2\frac{x_3^2+x_4^2}{2}\quad
  \text{and} \quad V(x_1,x_2,x_3,x_4)=-\frac{x_1^3}{3}+x_1x_3^2\, .
\end{equation}
Using the notation of the Poisson bracket -- which is defined by
$\{H\}f:=\{H,f\}=\sum_j\frac{\partial H}{\partial p_j}\frac{\partial
  f}{\partial q_j}-\frac{\partial H}{\partial q_j}\frac{\partial
  f}{\partial p_j}$ for any function $f(p,q)$ -- map \eqref{eq:4DFODO}
can be written as
\begin{equation}
  \vec{x}^{\prime}=T_{S}(\vec{x})=e^{\{H\}}e^{\{V\}}\vec{x}\, ,
\label{eq:pertflow}
\end{equation}
where $\vec{x}=(x_1,x_2,x_3,x_4)^{\rm T}\in\mathbb{R}^4$, $(^{\rm T})$
denotes the transpose of a matrix, and the exponential map is defined
as $e^{\{H\}}f=\sum_{n\geq 0}\frac{\{H\}^n}{n!}f$, with $
\{H\}^nf=\{H\}^{n-1}(\{ H\}f)$.

Assuming that the perturbation $V$ is small close to the origin, so
that $V=o(H)$, we can construct a {\em control map} whose generator
$F$ is small with respect to $V$ (i.e.~it satisfies $F=o(V)$).
Moreover, the {\em controlled map}
\begin{equation}
T_{ctrl}=e^{\{H\}}e^{\{V\}}e^{\{F\}}\, ,
\label{eq:controlled}
\end{equation}
is symplectic and conjugated to a map $T_*$, closer to $e^{\{H\}}$
than $T$ (for more details the reader is referred to
\cite{BC2010}). Assuming furthermore, that a non-resonant condition is
satisfied for the particular choice of the $q_x$ and $q_y$ values of
map \eqref{eq:4DFODO}, the generator $F$ of the control map
$e^{\{F\}}$ is
\begin{equation}
\label{eq:Ford2}
F=\frac{1}{2}\{V\}\mc{G}V +o(V^2)\, ,
\end{equation}
where $\mc{G}$ is a ``pseudo-inverse'' operator, which satisfies
$\mathcal{G} (1 - e^{-\mathcal{H}}) \mathcal{G} = \mathcal{G}$ (see
\cite{CVECP2005} for more details).

Even truncating the generator $F$ at order 2 and using
\begin{equation}
  F_2=\frac{1}{2}\{V\}\mc{G}V\, ,
\label{eq:F2}
\end{equation}
as an \textit{approximate generator}, the control map $e^{\{F_2\}}$
cannot, in general, be written in a closed analytic form. So, we
define a {\em truncated control map of order} $k$
\begin{equation}
  \label{eq:ctrlmapordk}
  C_k(F_2)=\sum_{l=0}^{k}\frac{\{F_2\}^l}{l!}\, ,
\end{equation}
and a {\em truncated controlled map of order} $k$:
\begin{equation}
  \label{eq:ctrlledmapordk}
  T_k(F_2)=e^{\{H\}}e^{\{V\}} C_k(F_2)=T_{S}\,C_k(F_2)\, .
\end{equation}

Following \cite{BC2010}, the Smaller Alignment Index (SALI)
method of chaos detection is used to determine the regular or chaotic nature
of orbits of map \eqref{eq:4DFODO} and its controlled version
\eqref{eq:controlled}. We note that an orbit is considered to escape
and collide with the accelerator's vacuum chamber if at some time $n$
$\sum_{i=1}^{4} x_i^2(n)>10$. For the evaluation of SALI, the
tangent of the studied map is computed and employed for following the evolution of two
initially linearly independent unit deviation vectors
$\hat{v}_{1}(0)$ and  $\hat{v}_{2}(0)$,  defining the index as
\begin{equation}\label{eq:SALI:2}
  \mbox{SALI}(n)=min \left\{\left\|\hat{v}_{1} {(n)}+\hat{v}_{2}
      {(n)}\right\|,\left\|\hat{v}_{1} {(n)}-\hat{v}_{2} {(n)}\right\|\right\}\, ,
\end{equation}
where $\| \cdot \|$ denotes the usual Euclidean norm and
$\hat{v}_{i}(n)=\frac{\vec{v}_{i}(n)}{\| \vec{v}_{i}(n)\|}$, $i=1,2$
are vectors of unit norm.

The behavior of SALI, and its generalization, the so-called Generalized
Alignment Index (GALI), has been studied in detail in
\cite{S01,SABV03,SABV04,SBA07}. According to these studies, in 4D maps
the SALI of chaotic orbits tends exponentially to zero as
$\mbox{SALI}(n)\propto e^{-(\sigma_1-\sigma_2)n}$ (with $\sigma_1$,
$\sigma_2$ being the two largest Lyapunov characteristic exponents of
the orbit), while it fluctuates around positive values,
i.e.~$\mbox{SALI}(n)\propto \mbox{const.}$, for regular orbits.  In
the case of 2D maps, the SALI tends to zero both for regular and
chaotic orbits, as $\mbox{SALI}(n)\propto 1/n^2$ and
$\mbox{SALI}(n)\propto e^{-\sigma_1 n}$, respectively. Thus, the
completely different behaviors of SALI for regular and chaotic orbits
allow us to clearly distinguish between the two case both for 4D and
2D maps.

In the following we consider an orbit to be chaotic
if $\mbox{SALI}(n) \leq 10^{-8}$ after $n$ iterations.  We note that in our studies we  typically have $\mbox{SALI}(n) \gtrsim 10^{-2}$ for regular orbits. Consequently the chosen threshold value ($\mbox{SALI}(n) \leq 10^{-8}$) safely guaranties the chaotic nature of orbits.

\section{The controlled 4D map}
\label{sec:4D_cont}

In \cite{BC2010} it was shown that the 4$^{th}$ order controlled map
$T_4(F_2)$, is a very good choice for the controlled system, since it
succeeded to considerably increase the DA of the accelerator, keeping
also the required CPU time for the evolution of large sets of initial
conditions, at acceptable levels.

The computed generator $F_2$ is a complicated function of both the
positions $x_1$, $x_3$ and the momenta $x_2$, $x_4$ of map
(\ref{eq:4DFODO}), and its specific expression is given in the
Appendix of \cite{BC2010}.

A simple approach for investigating the global dynamics of 4D
accelerator maps, used in \cite{BS2006,BC2010}, was the computation of
the maximal radius of a 4-dimensional hypersphere centered at
$(x_1,x_2,x_3,x_4)=(0,0,0,0)$, containing only regular orbits up to a finite but large number of iterations. This
approach provides a reliable indication of the size of the
DA. Application of this methodology in \cite{BC2010} showed that the
controlled map $T_4(F_2)$ has better stability properties than the
uncontrolled map (\ref{eq:4DFODO}). Nevertheless, one should keep in
mind that the center of these hyperspheres does not bear any
particular physical meaning, as it does not correspond to an
experimentally realizable beam configuration.

From a physical point of view it is more meaningful to study the
dynamics of the nominal orbit having $x_1(0)=x_3(0)=0$ for different
momenta $x_2$, $x_4$. For this purpose, we consider 4-dimensional
hyperspheres, centered at $(0,x_2,0,x_4)$ and compute for different
values of $x_2$, $x_4$, the radius $R$ of the largest hypersphere
which contains only regular orbits.

\begin{figure}[h]
\begin{center}
\psfig{file=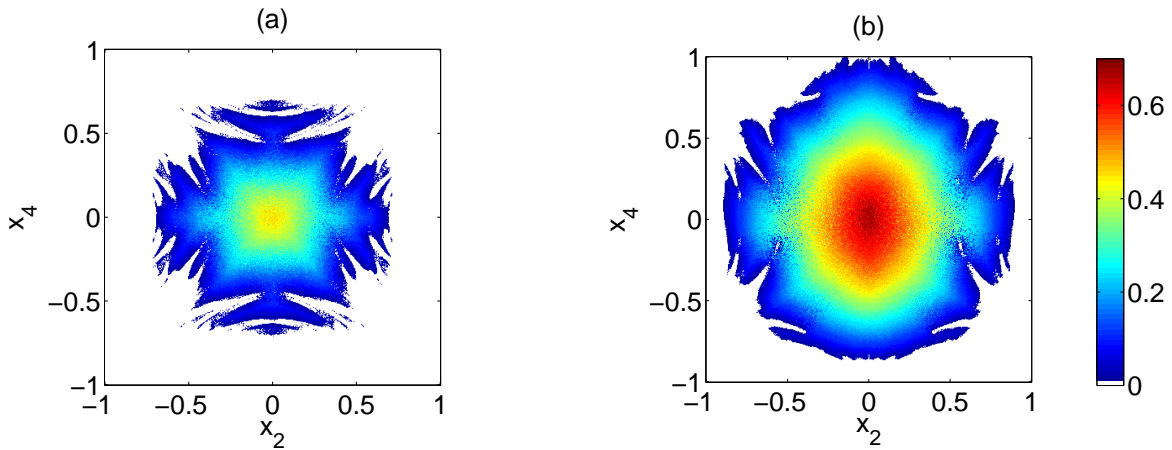,width=18cm} 
\end{center}
\caption{The $x_2$, $x_4$ coordinates of the centers $(0,x_2,0,x_4)$
  of the 4-dimensional hyperspheres containing only regular orbits for
  (a) the uncontrolled map (\ref{eq:4DFODO}), and (b) the $T_4(F_2)$
  map. Each point is colored according to the value of the hypersphere
  radius $R$ (with white color corresponding to $R=0$), and all orbits
  are evolved up to $10^4$ iterations.}
\label{mom}
\end{figure}

The outcome of this investigation is presented in Fig.~\ref{mom},
where the $x_2$, $x_4$ coordinates of the centers of the considered
hyperspheres are colored according to the value of radius $R$ for the
uncontrolled map (\ref{eq:4DFODO}) (Fig.~\ref{mom}(a)), and the
$T_4(F_2)$ map (Fig.~\ref{mom}(b)). For both maps the largest value of
$R$ is obtained for $x_2=x_4=0$ (i.e.~the value obtained in
\cite{BC2010}), while $R$ decreases as the hypershere's center is
moved away from $x_2=x_4=0$. This implies that large values of the
momenta introduce instabilities and chaotic behavior, which reduce the
size of the stability domain.

Nevertheless, the controlled map $T_4(F_2)$ succeeded to increase the
stability region, since $R >0$ for larger value of $x_2$ and $x_4$,
with respect to the original map. In addition, $R$ attains larger
values in the central region of the $(x_2,x_4)$ plane for the
controlled map $T_4(F_2)$, being $R\approx 0.6$, which suggests a
significant increase of the stability domain around the nominal orbit
of the accelerator.

As it has already being mentioned, the controlled map $T_4(F_2)$ has a
complicated functional form, depending both on spatial variables $x_1$,
$x_3$ and on the conjugate momenta $x_2$, $x_4$. From a practical
point of view, the addition of a control map (\ref{eq:ctrlmapordk})
can be thought as the addition of an appropriate magnetic element in
the ring. On the other hand, the vector potential of a Maxwellian magnetic field which
is transverse to the particle motion (i.e. without any longitudinal
dependence) is a function of these transverse positions and not momenta.
It is legitimate then to investigate the efficiency of a
modified control term, obtained by neglecting the dependence of $F_2$
on momenta. In practice, this means that we set $x_2=x_4=0$ in the
expression of $F_2$ obtained in \cite{BC2010}, getting
\begin{eqnarray}
\label{eq:gen}
F^{(0)}_2(x_1,x_3)&:=&F_2(x_1,0,x_3,0)=\nonumber\\ &=&\frac{1}{2}\, \left(
  x_1^{2}-x_3^{2} \right) \left\lbrace -\frac{1}{6}\,\csc \left( \frac{3
      \omega_1}{2} \right) \cos \left( \frac{\omega_1}{2} \right) \left[
    x_1^{2}-3\,x_3^{2}+ \left( 2\,x_1^{2}-6\,x_3^{2} \right) \cos \left( \omega_1
    \right) \right] + \nonumber \right. \\ && \left. +\frac{1}{6}\,\csc \left(
    \frac{3 \omega_1}{2} \right) x_1^{2}\sin \left( \frac{\omega_1}{2} \right)
  \sin \left( \omega_1 \right) -\frac{1}{4}\,{\frac {x_3^{2}\sin \left( \omega_2
      \right) }{\cos \left( \omega_2 \right) -\cos \left( \omega_1+\omega_2 \right)
    }} \right\rbrace + \nonumber\\ &&+\frac{1}{2}\,{\frac {x_1^{2}x_3^{2}\sin
    \left( \omega_2 \right) }{\cos \left( \omega_2 \right) -\cos \left(
      \omega_1+\omega_2 \right) }}.
 \end{eqnarray}
 This is a $4^{th}$ order polynomial in the positions which may be transformed
 with some variable rescaling to the vector potential of an octupole
 magnet with normal symmetry~\cite{W2007}
 \begin{equation}
 \label{eq:oct}
V_{oct} = b_4 Re\left[ \left( x_1 +i  x_3 \right)^4 \right]\, ,
\end{equation}
where $b_4$ denotes the multi-pole coefficient related to the magnet field strength.
 Since $F^{(0)}_2$ does not depend on the momenta, the control map
 $e^{\{F^{(0)}_2\}}$, can be explicitly computed. The action of this
 map on $x_2$ is given by
\begin{equation}
  e^{\{F_2^{(0)}\}}x_2=x_2+\{F_2^{(0)}\}x_2+\frac{\{F_2^{(0)}\}^2}{2!}x_2+\dots
  =x_2-\frac{\partial F_2^{(0)}}{\partial x_1}\, ,
\end{equation}
because $\{F_2^{(0)}\}x_2$ does not depend on the momenta and
consequently $\{F_2^{(0)}\}^mx_2=0$ for all integers $m\geq
2$. Similar relations hold also for the other variables,
i.e.~$\{F_2^{(0)}\}^mx_4=0$ for all integers $m\geq
2$ and $\{F_2^{(0)}\}x_1=\{F_2^{(0)}\}x_3=0$. Thus, no
truncation of the form (\ref{eq:ctrlmapordk}) is needed, and
consequently the simplified controlled map
\begin{equation}
\label{eq:momenta0}
T^{(0)}_{C}=e^{\{H\}}e^{\{V\}}e^{\{F^{(0)}_2\}}\, ,
\end{equation}
is symplectic by construction, as a composition of three explicitly
known symplectic maps.

\begin{figure}[h]
\begin{center}
\psfig{file=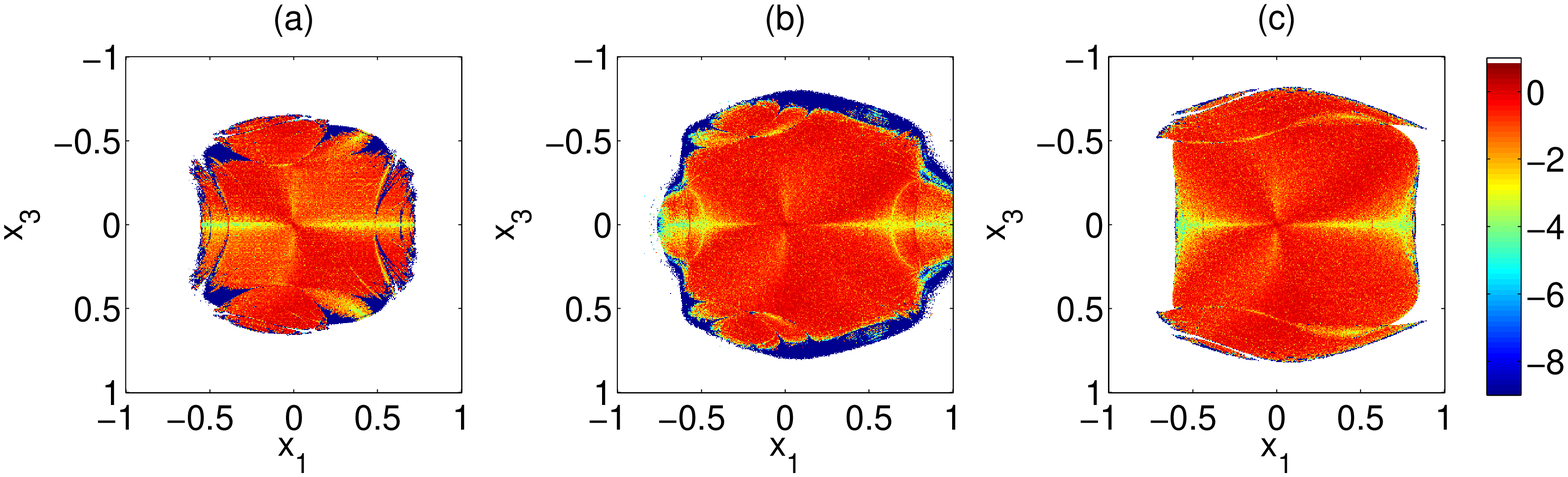,width=18cm}
\end{center}
\caption{Regions of different SALI values on the $(x_1,x_3)$ plane of
  (a) the uncontrolled map (\ref{eq:4DFODO}), (b) the $T_4(F_2)$
  controlled map (\ref{eq:ctrlledmapordk}), and (c) the $T^{(0)}_{C}$
  simplified controlled map (\ref{eq:momenta0}). $16000$ uniformly
  distributed initial conditions in the square
  $(x_1,x_3)\in[-1,1]\times [-1,1]$, $x_2(0)=x_4(0)=0$ are followed
  for $10^5$ iterations, and they are colored according to their final
  $\log_{10}\mbox{SALI}$ value. The white colored regions correspond
  to orbits that escape in less than $10^5$ iterations.}
\label{pp}
\end{figure}

Since map $T^{(0)}_{C}$ is not explicitly constructed by the
Hamiltonian Control Theory, it is interesting to check its performance
in controlling map (\ref{eq:4DFODO}) and increasing its DA. Following
\cite{BS2006,BC2010}, the SALI method is used to determine the regular
or chaotic nature of orbits of map $T^{(0)}_{C}$. In order to directly
compare these results with the ones obtained in previous studies,
the values of $\log_{10}\mbox{SALI}$ after $10^5$
iterations are plotted in Fig.~\ref{pp} , to describe the dynamics in the 2-dimensional subspace
$x_2(0)=x_4(0)=0$, for the uncontrolled map (\ref{eq:4DFODO})
(Fig.~\ref{pp}(a)), the $T_4(F_2)$ controlled map
(\ref{eq:ctrlledmapordk}) (Fig.~\ref{pp}(b)), and the simplified map
$T^{(0)}_{C}$ (\ref{eq:momenta0}) (Fig.~\ref{pp}(c)). Chaotic orbits
are characterized by small SALI values and are located in the blue
colored domains, regular orbits are colored red, while white regions
correspond to orbits that escape in less than $10^5$ iterations.
Fig.~\ref{pp} shows that both controlled maps, $T_4(F_2)$ and
$T^{(0)}_{C}$, increase the DA of the accelerator. It is remarkable
that the simplified controlled map $T^{(0)}_{C}$ not only increases
the region of non-escaping orbits, with respect to the original
system, but also decreases drastically the number of chaotic orbits.

\begin{figure}[h]
\begin{center}
\psfig{file=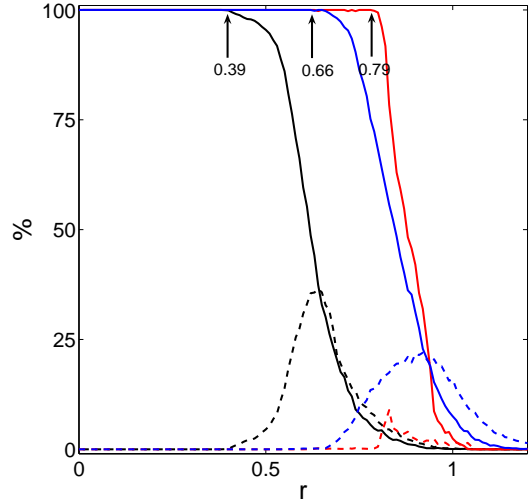,width=8cm}
\end{center}
\caption{The percentages of regular (solid curves) and chaotic (dashed
  curves) orbits after $n=10^5$ iterations of the original map
  (\ref{eq:4DFODO}) (black curves), the $T_4(F_2)$ controlled map
  (\ref{eq:ctrlledmapordk}) (blue curves), and the $T^{(0)}_{C}$
  simplified controlled map (\ref{eq:momenta0}) (red curves), in a
  4-dimensional hypersphere centered at $x_1=x_2=x_3=x_4=0$, as a
  function of the hypersphere radius $r$. The largest radii at which
  the percentage of regular orbits is $100\%$, are also marked.}
\label{da}
\end{figure}

In order to perform a more global investigation of the dynamics of
these maps we consider, as was done in \cite{BS2006,BC2010}, initial
conditions inside a 4-dimensional hypersphere centered at
$x_1=x_2=x_3=x_4=0$, and compute the percentages of regular and
chaotic orbits as a function of the hypersphere radius $r$
(Fig.~\ref{da}). From the red curves in Fig.~\ref{da},
it is observed that the $T^{(0)}_{C}$ map significantly increases the domain of regular
motion, not only with respect to the original map (black curves in
Fig.~\ref{da}), but also with respect to the controlled map $T_4(F_2)$
(blue curves in Fig.~\ref{da}). In addition, map $T^{(0)}_{C}$ has the
smallest percentage of chaotic orbits among the studied models, which
means that orbits of this map either escape or they are regular.

Thus, map $T^{(0)}_{C}$ (\ref{eq:momenta0}) controls the original
system (\ref{eq:4DFODO}) more efficiently than map $T_4(F_2)$
(\ref{eq:ctrlledmapordk}). Additional advantages of this map is its
simplicity, since it depends only on the spatial variables, and the
fact that it is symplectic by construction. Another feature, which
could be of major practical importance, is that the generator
$F^{(0)}_2$ (\ref{eq:gen}) of the simplified control map $T^{(0)}_{C}$
can be considered as the potential induced by a 
magnetic
element, since it depends only on spatial variables.

These results show that the simplified controlled map $T_C^{(0)}$ works better than the controlled map $T_4(F_2)$ which  is also
able to enlarge the DA with respect to $T_S$. Then, a natural task would be
to compare $T_C^{(0)}$ with  $T_{ctrl}$ or with
$T_{\infty}(F_2)$. We note that we were not able to find a closed
analytic form for the latter maps due to  the cumbersome involved summations. Thus, these comparisons remain open problems which we plan to address in the future. However, we remark  that the finite number of terms
involved in $T_4(F_2)$ possibly allows the attribution of a real magnetic
element to this map, while probably this will  not be feasible for maps
$T_{ctrl}$ and $T_{\infty}(F_2)$.

\section{Flat beams: the controlled 2D map}
\label{sec:2D_cont}

In many particle accelerators the beam is very flat, i.e. its vertical
extend is much smaller than the horizontal one. A simple, first,
approach in investigating the dynamics of a flat beam of the 4D map
(\ref{eq:4DFODO}) is to neglect the vertical coordinate $x_3$ and its
corresponding momentum $x_4$, and pass from the initial 4D map to the
2D map
\begin{equation}
  \label{eq:2DFODO}
  \left(
    \begin{matrix}
      x_1^{\prime}\\x_2^{\prime}
    \end{matrix}
\right)=\left(
  \begin{matrix}
    \cos \omega_1 & -\sin\omega_1 \\
    \sin \omega_1 & \cos\omega_1 \\
  \end{matrix}
\right)  \left(
    \begin{matrix}
      x_1\\x_2+x_1^2
    \end{matrix}
\right)=:T^{2D}\left(
    \begin{matrix}
      x_1\\x_2
    \end{matrix}
\right)\, ,
\end{equation}
whose dynamics was studied for example in \cite{BS2006b}.

\begin{figure}[h]
\begin{center}
\psfig{file=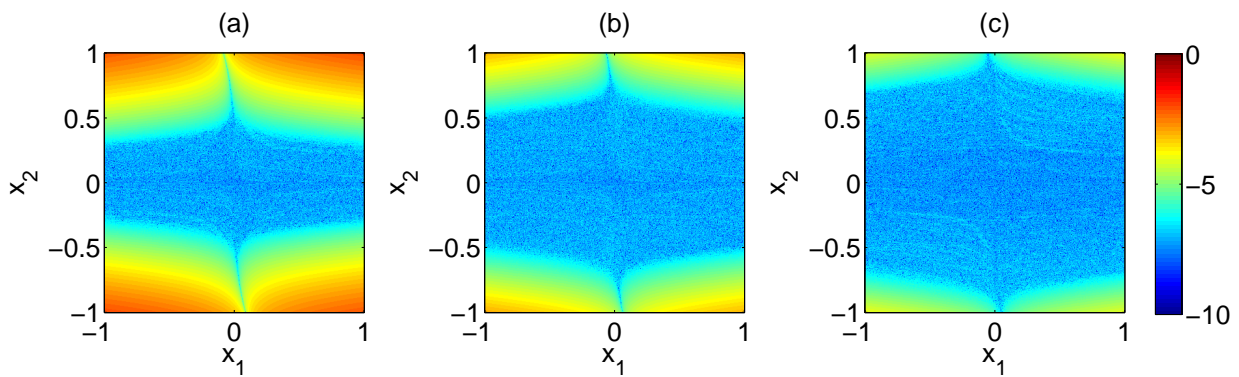,width=18cm}
\end{center}
\caption{Plot of $\log_{10}D_k$, where $D_k$ is the norm of matrix $A_k^T{J}A_k-{J}$, and
  $A_k$ is the Jacobian of the controlled map $T_k^{2D} (F_2^{2D})$,
  for $16000$ uniformly distributed values in the square $(x_1,x_2)\in
  [-1,1]\times[-1,1]$, for (a) $k=4$, (b) $k=6$, and (c) $k=8$. The
  percentages of orbits with $\log_{10}D_k < -6$ are $53\%$, $77\%$,
  and $94\%$ respectively for $k=2$, $4$, and $6$. The color scale
  corresponds to the value of $\log_{10}D_k$; the smaller the value of
  $\log_{10}D_k$ is (blue region), the closer the map is to a
  symplectic one.}
\label{symp}
\end{figure}

Following for this map, the procedure described in
Sect.~\ref{sec:methods}, truncated control $C_k^{2D} (F_2^{2D})$  and controlled  $T_k^{2D} (F_2^{2D})$
maps  of order $k$ are constructed, in analogy
to Eqs.~\eqref{eq:ctrlmapordk} and \eqref{eq:ctrlledmapordk},
respectively. As before, the controlled map is not necessarily symplectic.
Since a map is symplectic if its Jacobian matrix
$\mathbf{A}$ satisfies (in its definition domain) the equality $
\mathbf{A}^{\mathrm{T}}\mathbf{J}\mathbf{A}-\mathbf{J}=\mathbf{0} $
(with $\mathbf{J}=\left( \begin{array}{rc} \mathbf{0} & \mathbf{1} \\
    -\mathbf{1} & \mathbf{0}
  \end{array} \right) $ being  the standard symplectic constant matrix),
  the symplectic nature of $T_k^{2D} (F_2^{2D})$ can be checked
  by computing the norm $D_k$ of $\mathbf{A}^{\mathrm{T}}_k\mathbf{J}\mathbf{A}_k-\mathbf{J}$,
   for the Jacobian matrix $\mathbf{A}_k$ of the map. The results of this computation
are presented in Fig.~\ref{symp}, for orders $k=4$, $k=6$, and   $k=8$, in the region
$(x_1,x_2)\in [-1,1]\times[-1,1]$, and  show that
$T_k^{2D}(F_2^{2D})$ is a good approximation of a symplectic map for $k\geq 6$, since
$D_k\lesssim 10^{-6}$ for a large portion ($\gtrsim 77$ \%) of
variable values.  As expected, the larger the order $k$, the closer the map $T_k^{2D} (F_2^{2D})$ numerically approaches the symplectic condition.

\begin{figure}[h]
\begin{center}
\psfig{file=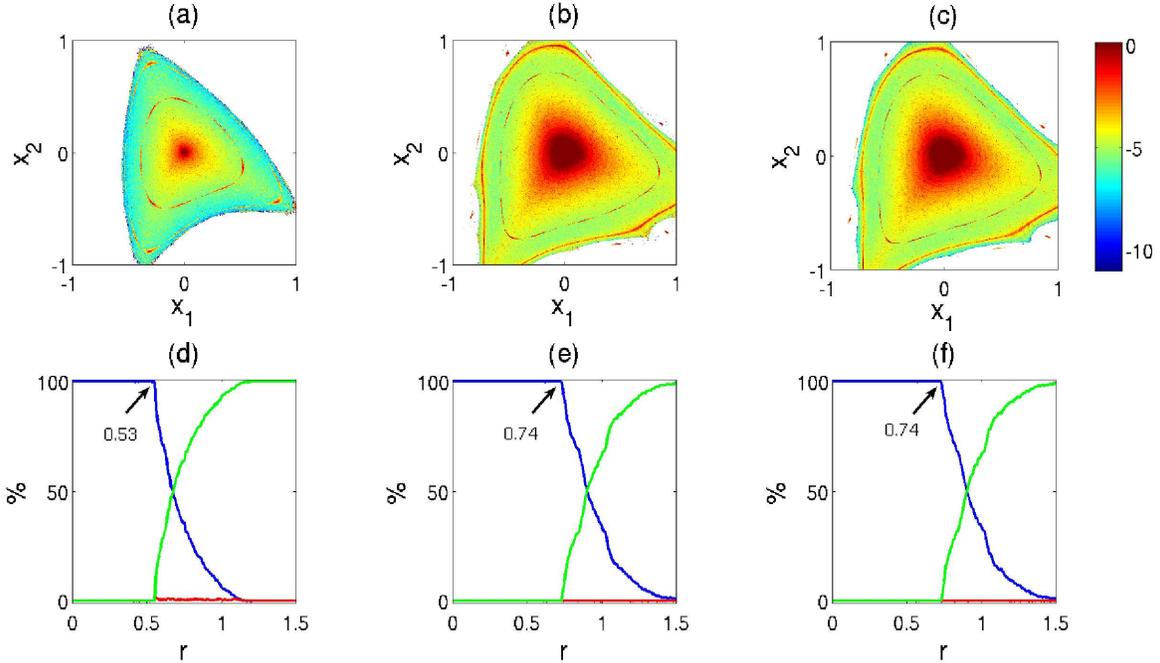,width=18cm}
\end{center}
\caption{Upper row: Regions of different SALI values on the
  $(x_1,x_2)$ plane of (a) the uncontrolled 2D map \eqref{eq:2DFODO},
  (b) the $T_6^{2D} (F_2^{2D})$ controlled map, and (c) the $T_8^{2D}
  (F_2^{2D})$ controlled map. $16000$ uniformly distributed initial
  conditions in the square $(x_1,x_2)\in[-1,1]\times [-1,1]$ are
  followed for $10^4$ iterations, and they are colored according to
  their final $\log_{10}\mbox{SALI}$ value. The white colored regions
  correspond to orbits that escape in less than $10^4$
  iterations. Lower row: The percentages of regular (blue curves),
  chaotic (red curves), and escaping (green curves) orbits after
  $n=10^4$ iterations of the maps of the upper row, in a circle
  centered at $x_1=x_2=0$, as a function of its radius $r$. The
  largest radii at which the percentage of regular orbits is $100\%$,
  are marked in panels (d)--(f). Panels in the same columns correspond
  to the same map.}
\label{2Dplein}
\end{figure}

In order to check whether the controlled map $T_k^{2D} (F_2^{2D})$
with $k\geq 4$ increases the DA of the flat beam, we use again the
SALI to determine the nature of orbits in the $(x_1,x_2)$ plane,
keeping in mind that SALI tends to zero both for regular and chaotic
orbits of 2D maps, but with time rates which allow the clear distinction
between the two cases. In particular, many initial conditions are tracked
for $10^4$ iterations in the $(x_1,x_2)$ plane, and colored
according to their final $\log_{10}\mbox{SALI}$ value, for the
$T^{2D}$ (Fig.~\ref{2Dplein}(a)), the $T_6^{2D} (F_2^{2D})$
(Fig.~\ref{2Dplein}(b)), and the $T_8^{2D} (F_2^{2D})$ map. Orbits
with $\mbox{SALI} \leq 10^{-10}$ are characterized as chaotic and are
colored in blue, while the remaining ones are regular. Similarly to
Fig.~\ref{pp} white regions correspond to escaping orbits. This distinction is based on the theoretical prediction that for regular orbits $\mbox{SALI}\propto 1/n^2=10^{-8}$ at $n=10^4$. In
Figs.~\ref{2Dplein}(d)--(f) the percentages of regular (blue curves),
chaotic (red curves), and escaping (green curves) orbits of these
three maps are plotted as a function of the radius $r$ of a circle
centered at the origin $x_1=x_2=0$.

In all models the number of chaotic orbits is practically negligible,
which means that orbits are either regular or escaping. The two
controlled maps succeed to increase the DA of the beam, because the
domain of non-escaping orbits increases
(Figs.~\ref{2Dplein}(a)--(c)). Although this domain does not have a
cyclical shape, the radius of the largest cycle containing only
regular orbits increases from $r\approx 0.53$ for the uncontrolled map
$T^{2D}$ \eqref{eq:2DFODO}, to $r\approx 0.74$ for both controlled
maps $T_6^{2D} (F_2^{2D})$ and $T_8^{2D} (F_2^{2D})$. Since both
controlled maps give essentially the same results, we conclude that
the $6^{th}$ order truncation of the controlled map is sufficient for
controlling the flat beam.

\section{Summary and discussion}

In this paper, the results of the application of control theory
to a simple 4D accelerator map~\cite{BC2010}, are being extended by using
simplified versions of the control maps. In the first case, the
momentum dependence of the control term is artificially removed,
 so as the control map is a function of the spacial coordinates and
 thus resembles the magnetic vector potential of a multi-pole magnet.
 In addition, this control map is symplectic by construction in contrast to the
 original one, thus avoiding the associated problems of having to choose
 an appropriate order of truncation in the Lie representation, for which the
 map satisfies numerically the symplectic condition.
 The efficiency of the simplified control map is remarkable, not only achieving
 the increase of the DA, but also shows a better performance
 as compared to the complete control map. The remaining important issue
 regarding the possibility to approximate in practice this magnetic field by
 a multi-pole magnet is still open, and will be addressed in a future study.
 In this respect, and based on the knowledge that control theory can be efficiently used to
 increase the DA of a toy model, the studies can be extended by investigating
 the possibility of imposing a control map with the functional form similar to Eq.~\eqref{eq:oct},
 and compute the associated multi-pole coefficients which achieve the best increase
 in the DA.

 In the second case, the same theory is applied to a 2D version of the map, modeling
 flat beams, as it is the case in electron and positron low emittance rings (i.e. with small beam
 sizes), where the vertical beam size is several orders of magnitude smaller than the horizontal
 one and thus the dynamics studies can be restricted to that plane. This 2D control map presents
  the same characteristics as the original map, i.e. increase of the DA, for a truncation order equal to 6.

  In all our studies, the SALI indicator was mainly used for showing the improvement of control
  in the DA. In future work, we plan to apply the frequency map analysis method (e.g. see~\cite{L95} and references therein) in this map in order to understand the dynamical details of this improvement with respect to resonance
  excitation and diffusion.

\nonumsection{Acknowledgments}

Numerical simulations were made on the local computing resources
(CLUSTER URBM--SYSDYN) at the University of Namur (FUNDP,
Belgium). Ch.~S.~was partly supported by the European research project
``Complex Matter'', funded by the GSRT of the Ministry Education of
Greece under the ERA-Network Complexity Program


\end{document}